\documentclass[journal]{IEEEtran}
\IEEEoverridecommandlockouts
\usepackage{cite}
\usepackage{amsmath,amssymb,amsfonts}
\usepackage{algorithmic}
\usepackage{graphicx}
\usepackage{textcomp}
\usepackage{subfig}
\usepackage{algorithm}
\usepackage{algorithmic}
\usepackage{amssymb}
\usepackage{math}
\usepackage{balance}
\usepackage{multirow}
\usepackage{booktabs}
\usepackage{enumitem}
\usepackage{gensymb}
\usepackage{array}
\usepackage{xcolor}
\usepackage[printonlyused,withpage]{acronym}
\usepackage{url}

\newcommand{\argmax}{\operatornamewithlimits{argmax}}

\newcommand{\inv}{^{\raisebox{.2ex}{$\scriptscriptstyle-1$}}}

\usepackage[final]{review}
\setcoverletter{coverletter-R1.tex}
\setrevision{1}

\graphicspath{{./Fig/}}

\def\BibTeX{{\rm B\kern-.05em{\sc i\kern-.025em b}\kern-.08em
    T\kern-.1667em\lower.7ex\hbox{E}\kern-.125emX}}

\acrodef{GMM}{Gaussian mixture model}
\acrodefplural{GMM}[GMMs]{Gaussian mixture models}
\acrodef{DOA}{direction of arrival}
\acrodefplural{DOA}[DOAs]{directions of arrival}   
\acrodef{pdf}{probability density function}
\acrodef{VEM}{variational expectation maximization}
\acrodef{TDOA}{time difference of arrival}
\acrodefplural{TDOA}[TDOAs]{time differences of arrival}
\acrodef{PHD}{probability hypothesis density}
\acrodef{MD}{missed detection}
\acrodefplural{MD}[MDs]{missed detections}
\acrodef{FA}{false alarm}
\acrodefplural{FA}[FAs]{false alarms}
\acrodef{MAE}{mean absolute error}
\acrodefplural{MAE}[MAEs]{mean absolute errors}
\acrodef{VAD}{voice activity detection}
\acrodef{i.i.d.}{independently and identically distributed}
\begin{document}

\title{Tracking Multiple Audio Sources with the\\ von Mises Distribution and Variational EM}
\author{
  Yutong Ban,$^1$ Xavier Alameda-Pineda,$^1$ \textit{Senior Member, IEEE},\\ Christine Evers,$^2$ \textit{Senior Member, IEEE}, and Radu Horaud$^1$
 \thanks{$^1$Y. Ban, X. Alameda-Pineda and R. Horaud are with Inria Grenoble Rh\^one-Alpes, Montbonnot Saint-Martin, France. E-mail: \texttt{first.last@inria.fr}}
 \thanks{$^2$C. Evers is with  Dept. Electrical and Electronic Engineering, Imperial College London, Exhibition Road, SW7 2AZ, UK. Email: \texttt{c.evers@imperial.ac.uk}}
 \thanks{This work was supported by the ERC Advanced Grant VHIA \#340113 and the UK EPSRC Fellowship grant no. EP/P001017/1.}
 \vspace{-3mm}
}


\maketitle
\begin{abstract}
In this paper we address the problem of simultaneously tracking several moving audio sources, namely the problem of estimating source trajectories from a sequence of observed features. We propose to use the von Mises distribution to model audio-source directions of arrival with circular random variables. 
This leads to a Bayesian filtering formulation which is intractable because of the combinatorial explosion of associating observed variables with latent variables, over time. We propose a variational approximation of the filtering distribution. We infer a variational expectation-maximization algorithm that is both computationally tractable and time efficient. 
We propose an audio-source birth method that favors smooth source trajectories and which is used both to initialize the number of active sources and to detect new sources. We perform experiments with the recently released LOCATA dataset comprising two moving sources and a moving microphone array mounted onto a robot.

%
\end{abstract}


\section{Introduction}
\label{sec:introduction}
We address the problem of tracking several moving audio sources. Audio tracking is useful for audio-source separation, spatial filtering, speaker diarization, speech enhancement and speech recognition, which in turn are essential methodologies, e.g. home assistants. Audio-source tracking is difficult because audio signals are adversely affected by noise, reverberation and interferences between acoustic signals. 

Single-source tracking methods are based on observing \acp{TDOA} between microphones. Since the mapping between \acp{TDOA} and the source locations  is non-linear, sequential Monte Carlo approaches are used, e.g. \cite{vermaak2001nonlinear,ward2003particle,zhong2014particle}. Alternatively, \acp{DOA} can be used. 
The problem is cast into a linear dynamic model, e.g.  \cite{bechler2003speaker}. In this case source directions should however be modeled as circular random variables, e.g. the wrapped Gaussian distribution \cite{traa2013wrapped}, or the von Mises distribution \cite{markovic2012bearing,evers2018doa}.

Multiple-source tracking is more challenging: (i)~the number of active sources is unknown and varies over time, (ii)~several \acp{DOA} need be detected, and (iii)~\ac{DOA}-to-source assignments must be estimated. An unknown number of sources was addressed using random finite sets~\cite{Mahler2003}. Since the \ac{pdf} is computationally intractable, its first-order approximation can be propagated in time using the \ac{PHD} filter \cite{Mahler2003,Vo2006}. In  \cite{ma2013} the \ac{PHD} filter was applied to audio recordings to track multiple sources from \ac{TDOA} estimates. In \cite{evers2018acoustic} the wrapped Gaussian distribution is incorporated within a \ac{PHD} filter. 
\addnote[newref]{1}
{
The von Mises-Fisher distribution was used in \cite{traa2014multiple} to build a factorial filter.
}
A mixture of von Mises distributions was combined with a \ac{PHD} filter in \cite{markovic2015mises}. \addnote[adhoc]{1}{The main drawback of \ac{PHD} filters is that explicit observation-to-source associations are not established. Instead, post-processing techniques are required for track labelling \cite{Lin2006track}}.

A variational approximation of the multiple target tracking was addressed in \cite{ba2016line}: bservation-to-target associations are discrete latent variables which are estimated with an \ac{VEM} solver. Moreover, the problem of tracking a varying number of targets is addressed via track-birth and track-death processes. The variational approximation of  \cite{ba2016line} was recently extended to track multiple speakers with audio \cite{li2018online} and audio-visual data \cite{ban2018variational}. 

This paper builds on  \cite{evers2018doa,ba2016line,li2018online} and proposes to use the von Mises distribution to model the \acp{DOA} of multiple acoustic sources with circular random variables. 
\addnote[novelty]{1}
{
The Bayesian filtering formulation for the multi-source tracking problem is intractable over time, due to the combinatorial nature of the unknown association between observed variables and latent variables. We propose a variational approximation of the filtering distribution. A novel mathematical framework is therefore proposed in order to deal with a mixture of von Mises distributions. The contribution of this paper is therefore a novel \ac{VEM} algorithm that is both computationally tractable and time efficient. 
}
Moreover, we propose an audio-source birth method that favors smooth source trajectories and which is used both to initialize the number of active sources and to detect new sources. We perform experiments with the recently released LOCATA dataset  \cite{LOCATA2018a} comprising audio recordings of two moving sources from a moving microphone array in a real acoustic environment.

The paper is organized as follows. Section~\ref{sec:model} describes the probabilistic model and Section~\ref{sec:variational} describes a variational approximation of the filtering distribution and the \ac{VEM} algorithm. Section~\ref{sec:birth_process} briefly describes the source birth method. Experiments and comparisons with other methods are described in Section~\ref{sec:experiments}. 
Supplemental materials (mathematical derivations, software and videos) are available online.\footnote{\url{https://team.inria.fr/perception/research/audiotrack-vonm/}}

\section{The Filtering Distribution}
\label{sec:model}
\addnote[definitions]{1}{
Let $N$ be the number of audio sources. Let $\yvect_t = \{y_{t1}, \dots y_{tm}, \dots y_{tM_t}\}$ be the set of $M_t$ \textit{observed} \acp{DOA}  at time step $t$. Let $\svect_{t} = \{s_{t1}, \dots, s_{tn}, \dots s_{tN}\}$ be the set of $N$ \textit{latent} \acp{DOA}, where $s_{tn}$ is the \ac{DOA} of source $n$ and time $t$. Observed and source \acp{DOA} are realizations of random circular variables $Y$ and $S$, respectively, in the interval $]-\pi,\pi]$, i.e. azimuth directions. Let $Z_{tm}$ be a discrete \textit{association} variable whose realizations take values in $\{0,1, \dots N\}$, i.e. $Z_{tm}=n$ means that observation $y_{tm}$ is assigned to source $n$ and $Z_{tm}=0$ means that the observation is ``clutter", hence assigned to none of the $N$ sources -- we refer to ``0" as a \textit{dummy} source. For convenience, we also use the notation $\zvect_t=\{z_{t1},\dots z_{tm}, \dots z_{tM_t}\}$.
}


Within a Bayesian model, multiple target tracking can be formulated as the estimation of the filtering distribution $p(\svect_t, \zvect_t | \yvect_{1:t})$, with the notation $\yvect_{1:t}=(\yvect_1, \dots \yvect_t)$. We assume that variables $s_{tn}$ follow a first-order Markov model, and that observations only depend on the current state and on the assignment variables. \addnote[dependencies]{1}{Moreover, we assume that the assignment variable does not depend on the previous observations.} Under these assumptions the posterior, or filtering, \ac{pdf} is given by:
\begin{align}
p(\svect_t, \zvect_t | \yvect_{1:t}) \propto  p(\yvect_t|   \zvect_t, \svect_t) p(\zvect_t) p(\svect_t | \yvect_{1:t-1}),
\label{eq:posterior_Bayes}
\end{align}
where $p(\yvect_t|   \zvect_t, \svect_t)$ is the observation likelihood, $p(\zvect_t)$ is the prior \ac{pdf} of the assignment variables and $p(\svect_t | \yvect_{1:t-1})$ is the predictive \ac{pdf} of the latent variables.

\subsubsection{Observation likelihood}
\label{subsec:audio-model}
Assuming that observed \acp{DOA} are \ac{i.i.d.}, the observation likelihood can be written as:
\begin{equation}
\label{eq:audio-independence}
p(\yvect_t| \zvect_t, \svect_t ) = \prod_{m=1}^{M_t} p (y_{tm} | \zvect_t, \svect_t).
\end{equation}
The likelihood that a \ac{DOA} corresponds to a source is modeled by a von Mises distribution \cite{evers2018doa}, whereas the likelihood that a \ac{DOA} corresponds to a dummy source (e.g. noise) is modeled by a uniform distribution:
\begin{align}
\label{eq:assign}
\!\! p (y_{tm} | Z_{tm} =n,  s_{tn} )=  
\begin{cases}
\mathcal{M} (y_{tm} ; s_{tn} , \kappa_{y} \omega_{tm}) &  n \neq 0 \\
\mathcal{U} (y_{tm}) & n=0
\end{cases},
\end{align}
where $\mathcal{M}(y\,; s , \kappa) =  (2\pi I_{0}(\kappa))^{-1} \exp  \lbrace \kappa \cos(y  - s) \rbrace$ denotes the von Mises distribution with mean $s$ and concentration $\kappa$, $I_{p}(\cdot)$ denotes the modified Bessel function of the first kind of order $p$, $\kappa_{y}$ denotes the concentration of audio observations, $\omega_{tm}\in [0,1]$ is a confidence associated with each observation, and $\mathcal{U}(y_{tm})=(2\pi)^{-1}$ denotes the uniform distribution along the support of the unit circle.
%

\subsubsection{Prior \ac{pdf} of the assignment variables}
Assuming that assignment variables are \ac{i.i.d.}, the joint prior \ac{pdf} is given by:
\begin{equation}
\label{eq:prior}
p(\zvect_{t}) = \prod_{m=1}^{M_t} p(Z_{tm}=n),
\end{equation}
and we denote with $\pi_n= p(Z_{tm}=n)$, $\sum_{n=0}^N \pi_n = 1$, the prior probability that source $n$ is associated with $y_{tm}$.

\subsubsection{Predictive \ac{pdf} of the latent variables}
The predictive \ac{pdf} extrapolates information inferred in the past to the current time step using a dynamic model for the source motion, i.e. \ac{DOA} rotation:
\begin{align}
\label{eq:predictive distribution}
p(\svect_t | \yvect_{1:t-1}) &= \int p(\svect_t | \svect_{t-1}) p( \svect_{t-1} | \yvect_{1:t-1}) d\svect_{t-1}.
\end{align}
where $p(s_t|s_{t-1})$ denotes the prior \ac{pdf} of the source motion and $p(s_{t-1}|y_{1:t-1})$ is the filtering \ac{pdf} at $t-1$. The sources are assumed to move independently, \addnote[doa-source]{1}{and each source (\ac{DOA}) follows a von Mises distribution}:
\begin{equation}
\label{eq:state_dynamic}
p(\svect_t | \svect_{t-1}) = \prod_{n=1}^{N} \mathcal{M}
(s_{tn}; s_{t-1,n}, \kappa_{d}),
\end{equation}
where $\kappa_{d}$ is the concentration of the state dynamics. $\Theta=\{\kappa_{y},\kappa_{d},\pi_0,\ldots,\pi_N\}$ denotes the set of model parameters.

As already mentioned in Section~\ref{sec:introduction}, the filtering distribution corresponds to a mixture model whose number of components grows exponentially along time, therefore solving (\ref{eq:posterior_Bayes}) directly is computationally intractable. Below we infer a variational approximation of~(\ref{eq:posterior_Bayes}) which drastically reduces the explosion of the number of mixture components; consequently, it leads to a computationally tractable algorithm.


%
%

\section{Variational Approximation and Algorithm}
\label{sec:variational}
Since solving~(\ref{eq:posterior_Bayes}) is computationally intractable, we propose to \addnote[var-approx]{1}{approximate the conditional independence between the latent and the assignment variables given all observations up to the current time step, t,}, more precisely
\begin{equation}
\label{eq:var-factor}
 p(\svect_t,\zvect_t|\yvect_{1:t})\approx q(\svect_t)q(\zvect_t).
\end{equation}
The proposed factorization leads to a \ac{VEM} algorithm~\cite{bishop2006pattern}, where the posterior distribution of the two variables are found by two variational E-steps:
\begin{align}
 q(\zvect_t)&\propto\exp\Big(\mathrm{E}_{q(\svect_t)} [\log p(\svect_t,\zvect_t|\yvect_{1:t})] \Big),\label{eq:variational-assignment}\\
 q(\svect_t)&\propto\exp\Big(\mathrm{E}_{q(\zvect_t)} [\log p(\svect_t,\zvect_t|\yvect_{1:t})] \Big),\label{eq:variational-state}
\end{align}
where \addnote[expectation]{1}{($E[\cdot]$ is the expectation operator).} The model parameters $\Theta$ are estimated by maximizing the expected complete-data log-likelihood:
\begin{equation}
Q(\Theta , \tilde{\Theta})=\mathrm{E}_{q(\svect_t)q(\zvect_t)}\Big[\log p(\yvect_t,\svect_t,\zvect_t|\yvect_{1:t-1},\Theta,\tilde{\Theta})\Big].
\label{eq:Qfunc}
\end{equation} 
where $\tilde{\Theta}$ are the old parameters. 
\addnote[independence]{1}
{
By combining the \ac{i.i.d.} assumption, i.e. \eqref{eq:audio-independence}, with the variational factorization \eqref{eq:var-factor}, we observe that the posterior \ac{pdf} of the assignment variables and the posterior \ac{pdf} of the latent variables can be factorized:
}
 \begin{equation} 
 \label{eq:var-separability}
  q(\zvect_t) = \prod_{m=1}^{M_t} q(z_{tm}), \quad q(\svect_t) = \prod_{n=1}^N q(s_{tn}),
 \end{equation}
%
and, therefore, the predictive \ac{pdf} is separable:
\begin{equation*}
 p(s_{tn}|\yvect_{1:t-1}) = \int p(s_{tn}|s_{t-1,n})p(s_{t-1,n}|\yvect_{1:t-1})\textrm{d}s_{t-1,n}.
\end{equation*}
Moreover, assuming that the filtering \ac{pdf} at $t-1$ follows a von Mises distribution, i.e. $q(s_{t-1,n}) = \mathcal{M}(s_{t-1,n};\mu_{t-1,n},\kappa_{t-1,n})$, then the predictive \ac{pdf} is approximately a von Mises distribution (see~\cite{evers2018doa}, \cite[(3.5.43)]{mardia2009directional}):
\begin{equation}
\label{eq:predictive-von-misses}
 p(s_{tn}|\yvect_{1:t-1})\approx\mathcal{M}(s_{tn};\mu_{t-1,n},\tilde{\kappa}_{t-1,n}),
\end{equation}
where the predicted concentration parameter, $\tilde{\kappa}_{t-1,n}$, is:
\begin{align}
    \label{eq:kappa_pred}
    \tilde{\kappa}_{t-1,n} = A^{-1}(A(\kappa_{t-1,n})A(\kappa_{d})),
\end{align}
and where $A(a) = I_{1}(a)/I_{0}(a)$, and 
$
A^{-1}(a) \approx (2a- a^{3})/(1-a^{2}).
$
%
Using (\ref{eq:variational-assignment}), (\ref{eq:variational-state}) and (\ref{eq:Qfunc}), the filtering distribution is therefore obtained by iterating through three steps, i.e. the E-S, E-Z and M steps, provided below (detailed mathematical derivations can be found in the appendices).

\subsubsection{E-S step}
\label{sec:post_s_dist_dev}
Inserting~\eqref{eq:posterior_Bayes} and~\eqref{eq:predictive-von-misses} in~\eqref{eq:variational-state}, $q(s_{tn})$ reduces to a von Mises distribution,
$\mathcal{M}(s_{tn}; \mu_{tn},\kappa_{tn})$. The mean $\mu_{tn}$ and concentration $\kappa_{tn}$ are given by:
\begin{align}
\mu	_{tn}& = \tan^{-1} \label{eq:mu_tn}\\
 &\left( \frac{ \kappa_{y}\sum_{m =1}^{M_t} \alpha_{tmn}\omega_{tm} \sin(y_{tm}) + \tilde{\kappa}_{t-1,n} \sin(\mu_{t-1,n})}
{\kappa_{y}\sum_{m =1}^{M_t} \alpha_{tmn}\omega_{tm} \cos(y_{tm}) + \tilde{\kappa}_{t-1,n} \cos(\mu_{t-1,n})} \right) 
\nonumber, \\[12px]
\kappa_{tn} &= \left( (\kappa_{y})^2\sum_{m =1}^{M_t} (\alpha_{tmn}\omega_{tm})^2 + \tilde{\kappa}_{t-1,n}^2 \right.\label{eq:kappa_tn}\\ 
&+ 2 (\kappa_{y})^{2}\sum_{m =1}^{M_t} \sum_{l =m+1}^{M_t} \alpha_{tmn}\omega_{tm}\alpha_{tln}w_{tl} \cos(y_{tm}-y_{tl}) \nonumber \\ 
&  \left. + 2 \kappa_{y}\tilde{\kappa}_{t-1,n}\sum_{m =1}^{M_t} (\alpha_{tmn}\omega_{tm} \cos(y_{tm} - \mu_{t-1,n})) 
\right)^{1/2}, \nonumber
\end{align}
where $\alpha_{tmn}=q(Z_{tm}=n)$ denotes the variational posterior probability of the assignment variables.
%
Therefore, the expressibility of the posterior distribution as a mixture of von Mises propagates over time, and only needs to be assumed at $t=1$. Please consult the supplementary materials for more details.

\subsubsection{E-Z step}
\label{sec:E-Z step}
By computing the expectation over $\svect_t$ in~(\ref{eq:variational-assignment}), the following expression is obtained:
\begin{equation}
\alpha_{tmn} = q(z_{tm} =n) = \frac{\pi_{n}\beta_{tmn}}{\sum_{l=0}^{N}\pi_{l}\beta_{tml}}\label{eq:alpha_tnm}
\end{equation}
where $\beta_{tmn}$ is given by (please consult the supplementary materials for a detailed derivation):
\begin{align*}
\label{eq:assign}
\!\!\!\beta_{tmn}  \!=\! \begin{cases}
\omega_{tm}\kappa_y A(\omega_{tm}\kappa_y) \cos(y_{tm}-\mu_{tn}) & \!n\neq0\\
1/(2\pi) & \!n=0,
\end{cases}
\end{align*}

\subsubsection{M step}
\label{sec:Mstep}
The parameter set $\Theta$ is evaluated by maximizing~(\ref{eq:Qfunc}). The priors \eqref{eq:prior} are obtained using the conventional update rule \cite{bishop2006pattern}: $\pi_n \propto \sum_{m=1}^{M_t} \alpha_{tnm}$. The concentration parameters, $\kappa_{y}$ and $\kappa_{d}$, are evaluated using gradient descent (please consult the supplementary materials).
Based on the E-S-step, E-Z-steo and M-step formulas above, the proposed \ac{VEM} algorithm iterates until convergence at each time step, in order to estimate the posterior distributions and to update the estimated model parameters.



\section{Audio-Source Birth Process}
\label{sec:birth_process}
We now describe in detail the proposed birth process which is essential to initialize the number of audio sources as well as to detect new sources at any time. The birth process gathers all the \acp{DOA} that were not assigned to a source, i.e. assigned to $n=0$, at current time $t$ as well over the $L$ previous times ($L=2$ in all our experiments). From this set of \acp{DOA} we build \ac{DOA}/observation sequences (one observation at each time $t$) and let $\hat{y}_{t-L:t}^j$ be such a sequence of \acp{DOA}, where $j$ is the sequence index. We consider the marginal likelihood:
\begin{equation}
    \label{eq:tau_int}
 \tau_j = p(\hat{y}_{t-L:t}^j) = \int p(\hat{y}_{t-L:t}^j,s_{t-L:t}) \textrm{d}s_{t-L:t}.
\end{equation}
Using \eqref{eq:predictive-von-misses} and the harmonic sum theorem, the integral (\ref{eq:tau_int}) becomes (please consult the supplementary materials):
\begin{equation}
\label{eq:tau_int_more}
 \tau_j = \prod_{l=0}^{L} \frac{I_0(\overline{\kappa}_{t-l}^j)}{2\pi I_0(\kappa_{y}\hat{\omega}_{t-l}^j) I_0(\hat{\kappa}_{t-l}^j)},
\end{equation}
where $\hat{\omega}_{t}$ is the confidence associated with $\hat{y}_t$. The concentration parameters, $\overline{\kappa}_{t-l}^j$ and $\hat{\kappa}_{t-l+1}^j$, depend on the observations and are recursively computed for each sequence $j$:
\begin{align*}
 &\overline{\kappa}_{t-l}^j = \sqrt{(\hat{\kappa}_{t-l}^j)^2 + (\kappa_{y}\hat{\omega}_{t-l}^j)^2 + \hat{\kappa}_{t-l}^j\kappa_{y}\hat{\omega}_{t-l}^j\cos(\hat{y}_{t-l}^j-\hat{\mu}_{t-l}^j)}, \\
 &\hat{\mu}_{t-l+1}^j =  \tan \inv \left( \frac{\hat{\kappa}_{t-l}^j\sin(\hat{\mu}_{t-l}^j) + \kappa_{y}\hat{\omega}_{t-l}^j\sin(\hat{y}_{t-l}^j)}{\hat{\kappa}_{t-l}^j\cos(\hat{\mu}_{t-l}^j) + \kappa_{y}\hat{\omega}_{t-l}^j\cos(\hat{y}_{t-l}^j)} \right),\\
 &\hat{\kappa}_{t-l+1}^j = A^{-1}(A(\tilde{\kappa}_{t-l}^j)A(\kappa_{d})).
\end{align*}
The sequence $j^\ast$ with the maximal marginal likelihood \eqref{eq:tau_int_more}, namely $j^\ast = \argmax_j  (\tau_j ) $, is supposed to be generated from a not yet known audio source only if $ \tau_{j^\ast} $ is larger than a threshold $\tau_0$: a new source $\tilde{n}$ is created in this case and $q(s_{t \tilde{n}}) = \mathcal{M}(s_{t \tilde{n}};\hat{\mu}_{t j^\ast},\hat{\kappa}_{t j^\ast})$. 
\addnote[death]{1}
{
We note that, in practice, a source may become silent. In this case, the source is no longer associated with observations, and the proposed tracking algorithm relies solely on the source dynamics. If a source is silent for a long time the algorithm loses track of that source. If, after a while, the source becomes active again, a new track is initialized.
}

\section{Experimental Evaluation}\label{sec:experiments}
The proposed method was evaluated using the audio recordings from Task~6 of the IEEE-AASP LOCATA\footnote{\url{https://locata.lms.tf.fau.de/}} challenge development dataset \cite{LOCATA2018a}, which involves multiple moving sound sources, i.e. speakers, and a microphone array mounted onto the head of a biped humanoid robot. The LOCATA dataset  consists of real-world recordings with ground-truth source locations provided by an optical tracking system. The size of the recording room is $7.1 \times 9.8 \times 3$~m, with $T_{60} \approx 0.55$~s. Task~6 contains three sequences of a total duration of $188.4$~s and two moving speakers. In our experiments we used four coplanar microphones, namely \#5, \#8, \#11, and \#12. The online sound-source localization method~\cite{li2018online} was used to provide \ac{DOA} estimates at each STFT frame, using a Hamming window of length 16~ms, with 8~ms shifts. \addnote[peakdetection]{1}{The approach in ~\cite{li2018online} requires a threshold, set to $0.3$ in our case, to select the number of significant active source, observed source \acp{DOA}, and the associated confidence values (see \cite{li2017taslp,li2018online})}. \addnote[birththresh]{1}{The birth  threshold, $\tau_0$, is set to 0.5 (Section~\ref{sec:birth_process}).} 
  
To evaluate the method quantitatively, the estimated source trajectories are compared with the ground-truth trajectories over audio-active frames. Ground-truth audio-active frames are obtained using the \ac{VAD} method of~\cite{li2016voice}. The permutation problem between the detected trajectories and the ground-truth trajectories is solved by means of a greedy gating algorithm: the error between all possible pairs of estimated and ground-truth trajectories is evaluated. Minimum-error pairs are selected for further comparison. A \ac{DOA} estimate that is 15$^{\circ}$ away from the ground-truth is treated as a false alarm detection. Sources that are not associated with a trajectory correspond to \acp{MD}. For performance evaluation, the percentage of \acp{MD} and \acp{FA} are evaluated over voice-active frames. The \ac{MAE} the error between ground-truth \acp{DOA} and estimated \ac{DOA} over all the active frames of all the speakers.

The observation-to-source assignment posteriors and the \acp{DOA} confidence weights are used to estimate voice-active frames:
\begin{equation}
\sum_{t' = t-D}^{t} \sum_{m =1}^{M_t} \alpha_{t'mn} \omega_{t'm} {{active \atop >}\atop{<\atop silent}} \delta
\end{equation}  
where $D=2$ and $\delta=0.025$ is a \ac{VAD} threshold. Once an active source is detected, we output its trajectory.
 
\begin{table}
\begin{center}
\begin{tabular}{lccc}
\toprule
Method & \ac{MD} (\%) & \ac{FA} (\%) & \ac{MAE} (\degree)\\
\midrule
vM-PHD~\cite{markovic2015mises} & 33.4 & 9.5 & 4.5 \\
\midrule
GM-ZO~\cite{li2018online} & 27.0 & 10.8 & 4.7 \\
\midrule
GM-FO~\cite{li2018online} & \textbf{22.3} & 6.3 & 3.2 \\
\midrule
vM-VEM (proposed) & 23.9  & \textbf{5.9} & \textbf{2.6}\\
\bottomrule
\end{tabular}
\end{center}
\caption{Method evaluation with the LOCATA dataset.}
\label{tab:quantitative results}
\end{table}

\begin{figure}[t]
\begin{center}
\begin{tabular}{cc}
\includegraphics[scale=0.29]{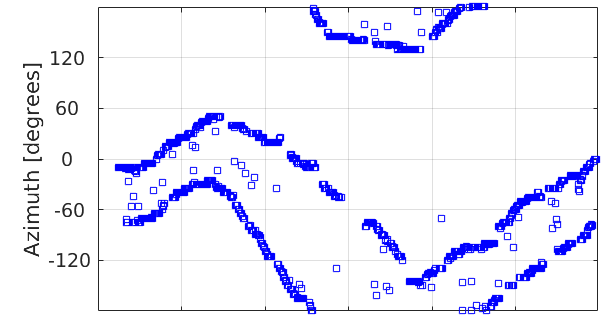}\hspace{-0.4cm} & 
\includegraphics[scale=0.29]{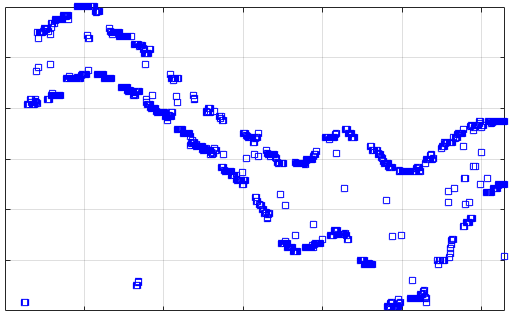} \\
\includegraphics[scale=0.29]{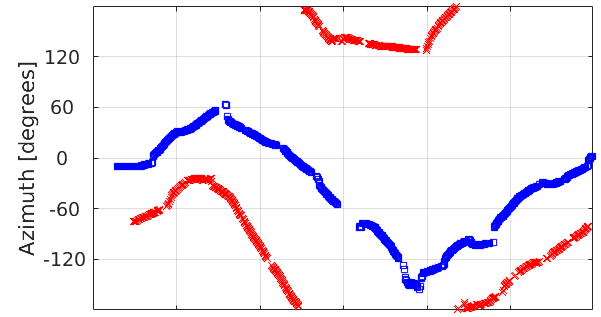}\hspace{-0.4cm} & 
\includegraphics[scale=0.29]{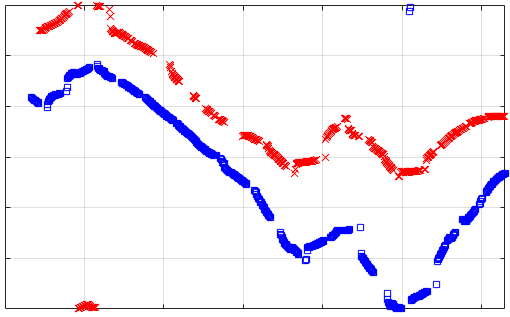} \\
\includegraphics[scale=0.29]{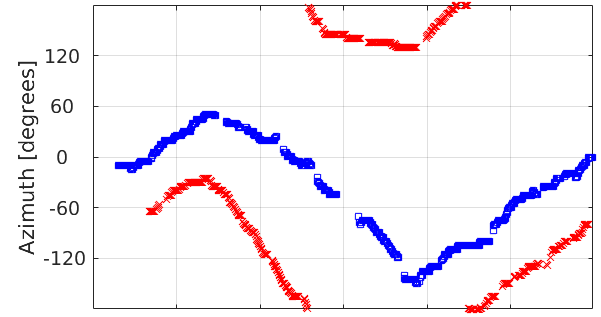}\hspace{-0.4cm} & 
\includegraphics[scale=0.29]{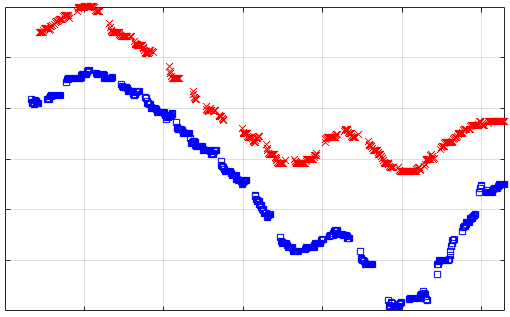} \\
\includegraphics[scale=0.29]{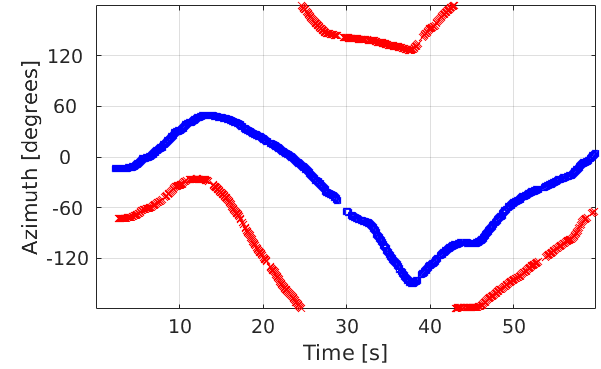}\hspace{-0.4cm} & 
\includegraphics[scale=0.29]{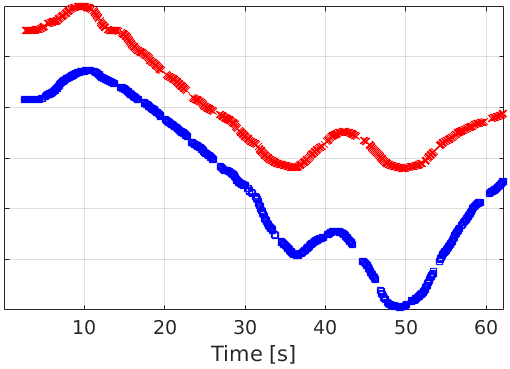} \\
\end{tabular}
\end{center}
\caption{Results obtained with recordings \#1 (left) and \#2 (right) from Task 6 of the LOCATA dataset. Top-to-down: vM-PHD~\cite{markovic2015mises}, GM-FO~\cite{li2018online}, vM-VEM (proposed) and ground-truth trajectories. Different colors represent different audio sources. Note that vM-PHD is unable to associate sources with trajectories.}
\label{fig:results}
\end{figure}

The \acp{MAE}, \acp{MD} and \acp{FA} values, averaged over all recordings, are summarized in Table~\ref{tab:quantitative results}.   We compared the proposed von Mises \ac{VEM} algorithm (vM-VEM) with three multi-speaker trackers: the von Mises \ac{PHD} filter (vM-PHD)~\cite{markovic2015mises} and two versions  the multiple speaker tracker of~\cite{li2018online} based on Gaussians models (GM). \cite{li2018online} uses a first-order dynamic model whose effect is to smooth the estimated trajectories. We compared with both first-order (GM-FO) and zero-order (GM-ZO) dynamics. The proposed vM-VEM tracker yields the lowest false alarm (FA) rate of $5.9 \%$ and \ac{MAE} of $2.6$, and the second lowest \ac{MD} rate of $23.9\%$. The GM-FO variant of  \cite{li2018online} yields an \ac{MD} rate of $22.3\%$ since it uses velocity information to smooth the trajectories. This illustrates the advantage of the von-Mises distribution to model directional data (\ac{DOA}). The proposed von-Mises model uses a zero-order dynamics; nevertheless it achieves performance comparable with the Gaussian model that uses first-order dynamics.

%

The results for recordings \#1 and \#2 in Task 6 are shown in Fig.~\ref{fig:results}, \addnote[sampling]{1}
{using a sampling rate of 12 Hz for plotting}.
 Note that the PHD-based filter method \cite{markovic2015mises} has two  caveats. First, 
observation-to-source assignments cannot be estimated (unless a post-processing step is performed), and second, the estimated source trajectories are not smooth. This stays in contrast with the proposed method which explicitly represents assignments with discrete latent variables and estimates them iteratively with \ac{VEM}.  Moreover, the proposed method yields smooth trajectories similar with those estimated by \cite{li2018online} and quite close to the ground truth.
%
%

\section{Conclusion}
\label{sec:conclusion}
We proposed a multiple audio-source tracking method using the von Mises distribution and we inferred a tractable solver based on a variational approximation of the posterior filtering distribution. Unlike the wrapped Gaussian distribution, the von Mises distribution explicitly models the circular variables associated with audio-source localization and tracking based on source \acp{DOA}. Using the recently released LOCATA dataset, we empirically showed that the proposed method compares favorably with two recent methods.

\appendices
\section{Derivation of the E-S step}
\label{sec:appendix-e-s-step}
In order to obtain the formulae for the E-S step, we start from its definition in~(9): 
\begin{equation}
     q(\svect_t)\propto\exp\Big(\mathrm{E}_{q(\zvect_t)}\log p(\svect_t,\zvect_t|\yvect_{1:t}) \Big).
\end{equation}
We now use the decomposition in~(1) to write: 
\begin{equation}
     q(\svect_t)\propto\exp\Big(\mathrm{E}_{q(\zvect_t)}\log p(\yvect_t|\svect_t,\zvect_t) \Big) p(\svect_t|\yvect_{1:t-1}).
     \label{eq:variational-state-app}
\end{equation}
Let us now develop the expectation:
\begin{align*}
& \mathrm{E}_{q(\zvect_t)}\log p(\yvect_t|\svect_t,\zvect_t) \\
& = \mathrm{E}_{q(\zvect_t)} \sum_{m=1}^{M_t} \log p(y_{tm}|\svect_t,z_{tm}) \\
& = \sum_{m=1}^{M_t} \mathrm{E}_{q(z_{tm})} \log p(y_{tm}|\svect_t,z_{tm}) \\
& = \sum_{m=1}^{M_t} \sum_{n=0}^N q(z_{tm}=n) \log p(y_{tm}|\svect_t,z_{tm}=n) \\
& = \sum_{m=1}^{M_t} \sum_{n=0}^N \alpha_{tnm} \log p(y_{tm}|s_{tn},z_{tm}=n) \\
& = \sum_{m=1}^{M_t} \sum_{n=0}^N \alpha_{tnm} \log {\cal M}(y_{tm};s_{tn},\omega_{tm}\kappa_y) \\
& \stackrel{\svect_{t}}{=} \sum_{m=1}^{M_t} \sum_{n=0}^N \alpha_{tnm}\omega_{tm}\kappa_y\cos(y_{tm}-s_{tn}),
\end{align*}
where $\stackrel{\svect_{t}}{=}$ denotes the equality up to an additive constant that does \textit{not} depend on $\svect_{t}$. Such a constant would become a multiplicative constant after the exponentiation in~(\ref{eq:variational-state-app}), 
and therefore can be ignored.

By replacing the developed expectation together with~(12) we obtain: 
\begin{align*}
     q(\svect_t)\propto & \exp\Big(\sum_{m=1}^{M_t} \sum_{n=0}^N \alpha_{tnm}\omega_{tm}\kappa_y\cos(y_{tm}-s_{tn}) \Big)\\
     & \prod_{n=0}^N {\cal M}(\svect_{tn};\mu_{t-1,n},\tilde{\kappa}_{t-1,n}),
\end{align*}
which can be rewritten as:
\begin{align}
     q(\svect_t)\propto  \prod_{n=0}^N \exp\Big( & \sum_{m=1}^{M_t} \alpha_{tnm}\omega_{tm}\kappa_y\cos(y_{tm}-s_{tn}) \\
     &  + \tilde{\kappa}_{t-1,n}\cos(s_{tn}-\mu_{t-1,n})\Big).
     \label{eq:qs-factorize}
\end{align}

\eqref{eq:qs-factorize} is important since it demonstrates that the \emph{a posteriori} pdf of $\svect_t$ is separable on $n$ and therefore independent for each speaker. In addition, it allows us to rewrite the \emph{a posteriori} pdf for each speaker, i.e.,\ of $s_{tn}$ as a von Mises distribution by using the harmonic addition theorem, thus obtaining
\begin{equation}
    q(\svect_t) = \prod_{n=0}^N q(s_{tn}) = \prod_{n=0}^N {\cal M}(s_{tn};\mu_{tn},\kappa_{tn}),
\end{equation}
with $\mu_{tn}$ and $\kappa_{tn}$ defined as in~(14) and (15). 

\section{Derivation of the E-Z step}
\label{sec:appendix-e-z-step}
Similarly to the previous section, and in order to obtain the closed-form solution of the E-Z step, we start from its definition in~(8): 
\begin{equation}
     q(\zvect_t)\propto\exp\Big(\mathrm{E}_{q(\svect_t)}\log p(\svect_t,\zvect_t|\yvect_{1:t}) \Big),
\end{equation}
and we use the decomposition in (1),
\begin{equation}
     q(\zvect_t)\propto\exp\Big(\mathrm{E}_{q(\svect_t)}\log p(\yvect_t|\svect_t,\zvect_t) \Big) p(\zvect_t).
\end{equation}

Since both the observation likelihood and the prior distribution are separable on $z_{tm}$, we can write:
\begin{equation}
     q(\zvect_t)\propto\prod_{m=1}^{M_t}\exp\Big(\mathrm{E}_{q(\svect_t)}\log p(y_{tm}|\svect_t,z_{tm}) \Big) p(z_{tm}),
\end{equation}
proving that the \emph{a posteriori} pdf is also separable on $m$.

We can thus analyze the posterior of each $z_{tm}$ separately, by computing $q(z_{tm}=n)$:
\begin{align*}
     q(z_{tm}=n)\propto & \exp\Big(\mathrm{E}_{q(\svect_t)}\log p(y_{tm}|\svect_t,z_{tm}=n) \Big) p(z_{tm}=n)
\end{align*}

Let us first compute the expectation for $n\neq 0$:
\begin{align*}
   & \mathrm{E}_{q(\svect_t)}\log p(y_{tm}|\svect_t,z_{tm}=n)\\
   & = \mathrm{E}_{q(s_{tn})}\log p(y_{tm}|s_{tn},z_{tm}=n) \\
   & = \mathrm{E}_{q(s_{tn})}\log {\cal M}(y_{tm};s_{tn},\omega_{tm}\kappa_y) \\
   & \stackrel{z_{tm}}{=} \int_{0}^{2\pi} q(s_{tn}) \omega_{tm}\kappa_y\cos(y_{tm}-s_{tn})\textrm{d}s_{tn} \\
   & = \frac{\omega_{tm}\kappa_y}{2\pi I_0(\omega_{tm}\kappa_y)} \int_{0}^{2\pi} \!\!\!\exp\Big(\!\cos(s_{tn}-\mu_{tn})\Big)\cos(s_{tn}-y_{tm})\textrm{d}s_{tn} \\
   & = \omega_{tm}\kappa_y A(\omega_{tm}\kappa_y) \cos(y_{tm}-\mu_{tn}),
\end{align*}
where for the last line we used the following variable change $\bar{s}=s_{tn}-\mu_{tn}$ and the definition of $I_1$ and $A$.

The case $n=0$ is even easier since the observation distribution is a uniform: $\mathrm{E}_{q(s_{tn})}\log p(y_{tm}|s_{tn},z_{tm}=n)=\mathrm{E}_{q(s_{tn})} -\log 2\pi = -\log(2\pi)$.

By using the fact that the prior distribution on $z_{tm}$ is denoted by $p(z_{tm}=n) = \pi_n$, we can now write the a posteriori distribution as $q(z_{tm}=n)  \propto \pi_n \beta_{tmn}$ with:
\begin{align*}
    \beta_{tmn} &= \left\{\begin{array}{ll}
    \omega_{tm}\kappa_y A(\omega_{tm}\kappa_y) \cos(y_{tm}-\mu_{tn}) & n\neq 0 \\
    1/2\pi & n=0
    \end{array}\right. ,
\end{align*}
thus leading to the results in~(16) and (3). 

\section{Derivation of the M step}
\label{sec:appendix-m-step}
In order to derive the M step, we need first to compute the $Q$ function in~(10), 
\begin{align*}
    Q(\Theta,\tilde{\Theta}) &= \mathrm{E}_{q(\svect_t)q(\zvect_t)}\Big\{\log p(\yvect_t,\svect_t,\zvect_t|\yvect_{1:t-1},\Theta)\Big\} \\
    &= \mathrm{E}_{q(\svect_t)q(\zvect_t)}\Big\{ \underbrace{\log p(\yvect_t|\svect_t,\zvect_t,\Theta)}_{\kappa_y} + \\
    &= + \underbrace{\log p(\zvect_t|\Theta)}_{\pi_n's} + \underbrace{\log p(\svect_t|\yvect_{1:t-1},\Theta)}_{\kappa_d} \Big\},
\end{align*}
where each parameter is show below the corresponding term of the $Q$ function. Let us develop each term separately.

\subsection{Optimizing $\kappa_y$}
\begin{align*}
    &Q_{\kappa_y} = \mathrm{E}_{q(\svect_t)q(\zvect_t)}\Big\{\log\prod_{m=1}^{M_t} p(y_{tm}|\svect_t,z_{tm})\Big\}\\
    &=\sum_{m=1}^{M_t} \mathrm{E}_{q(\svect_t)q(z_{tm})}\Big\{\log p(y_{tm}|\svect_t,z_{tm})\Big\} \\
    &=\sum_{m=1}^{M_t} \mathrm{E}_{q(\svect_t)}\sum_{n=0}^N\alpha_{tmn}\Big\{\log p(y_{tm}|\svect_t,z_{tm}=n)\Big\} \\
    &=\sum_{m=1}^{M_t} \sum_{n=0}^N\alpha_{tmn}\mathrm{E}_{q(s_{tn})}\Big\{\log {\cal M}(y_{tm};s_{tn},\omega_{tm}\kappa_y)\Big\} \\
    &=\sum_{m=1}^{M_t} \sum_{n=0}^N\alpha_{tmn}\int_{0}^{2\pi}\!\!\!\!\!\!q(s_{tn}) (\omega_{tm}\kappa_y\cos(y_{tm}-s_{tn}) \\ & \ \ \ \ \ \ \ \ \ \ \ \ \ \ \ \ \ \ \ \ \ \ \ \ \ \ \ \ \ \ \ \ \ - \log(I_0(\omega_{tm}\kappa_y)))\textrm{d}s_{tn} \\
    &=\sum_{m=1}^{M_t} \sum_{n=0}^N\alpha_{tmn}\Big(\omega_{tm}\kappa_y\cos(y_{tm}-\mu_{tn})A(\kappa_{tn}) - \log(I_0(\omega_{tm}\kappa_y))\Big),
\end{align*}
and by taking the derivative with respect to $\kappa_y$ we obtain:
\begin{equation*}
\frac{\partial Q}{\partial \kappa_y} = \sum_{m=1}^{M_t} \sum_{n=0}^N\alpha_{tmn}\omega_{tm}\Big(\cos(y_{tm}-\mu_{tn})A(\kappa_{tn}) - A(\omega_{tm}\kappa_y)\Big),
\end{equation*}
which corresponds to what was announced in the manuscript.

\subsection{Optimizing $\pi_n$'s}
\begin{align*}
    &Q_{\pi_n} = \mathrm{E}_{q(\svect_t)q(\zvect_t)}\Big\{\log\prod_{m=1}^{M_t} p(z_{tm})\Big\}\\
    & = \sum_{m=1}^{M_t}\mathrm{E}_{q(z_{tm})}\Big\{\log p(z_{tm})\Big\}\\
    & = \sum_{m=1}^{M_t}\sum_{n=0}^N\alpha_{tmn}\Big\{\log p(z_{tm}=n)\Big\}\\
    & = \sum_{m=1}^{M_t}\sum_{n=0}^N\alpha_{tmn}\Big\{\log \pi_n\Big\}
\end{align*}

This is the same formulae that is correct for any mixture model, and therefore the solution is standard and corresponds to the one reported in the manuscript.

\subsection{Optimizing $\kappa_d$}
\begin{align*}
    &Q_{\kappa_d} = \mathrm{E}_{q(\svect_t)q(\zvect_t)}\Big\{\log\prod_{n=1}^{N} p(s_{tn}|\yvect_{1:t-1})\Big\}\\
    &=\sum_{n=1}^N \mathrm{E}_{q(s_{tn})}\Big\{\log {\cal M}(s_{tn};\mu_{t-1,n},\tilde{\kappa}_{t-1,n})\Big\}\\
    &=\sum_{n=1}^N \mathrm{E}_{q(s_{tn})}\Big\{-\log I_0(\tilde{\kappa}_{t-1,n}) + \tilde{\kappa}_{t-1,n}\cos(s_{tn}-\mu_{t-1,n})\Big\}\\
    &=\sum_{n=1}^N -\log I_0(\tilde{\kappa}_{t-1,n}) + \tilde{\kappa}_{t-1,n}\cos(\mu_{tn}-\mu_{t-1,n})A(\kappa_{tn}),
\end{align*}
where the dependency on $\kappa_d$ is implicit in $\tilde{\kappa}_{t-1,n}=A^{-1}(A(\kappa_{t-1,n})A(\kappa_d))$.

By taking the derivative with respect to $\kappa_d$ we obtain:
\begin{equation*}
    \frac{\partial Q}{\partial \kappa_d} = \sum_{n=1}^N \Big(A(\kappa_{tn})\cos(\mu_{tn}-\mu_{t-1,n}) - A(\tilde{\kappa}_{t-1,n})\Big)\frac{\partial \tilde{\kappa}_{t-1,n}}{\partial \kappa_d}
\end{equation*}
with
\begin{align*}
    \frac{\partial \tilde{\kappa}_{t-1,n}}{\partial \kappa_d} &= \tilde{A}(A(\kappa_{t-1,n})A(\kappa_d))A(\kappa_{t-1,n})\frac{I_2(\kappa_{d})I_0(\kappa_{d})-I_1^2(\kappa_{d})}{I_0^2(\kappa_{d})},
\end{align*}
where $\tilde{A}(a)=\textrm{d}A^{-1}(a)/\textrm{d}a = (2-a^2+a^4)/(1-a^2)^2$.

By denoting the previous derivative as $B(\kappa_d)=\frac{\partial \tilde{\kappa}_{t-1,n}}{\partial \kappa_d}$, we obtain the expression in the manuscript.

\section{Derivation of the birth probability}
\label{sec:appendix-birth}

In this section we derive the expression for $\tau_j$ by computing the integral~(17). 
Using the probabilistic model defined, we can write (the index $j$ is omitted):
\begin{align*}
 \int& p(\hat{y}_{t-L:t},s_{t-L:t}) \textrm{d}s_{t-L:t}\\
  &= \int\prod_{\tau=-L}^0 p(\hat{y}_{t+\tau}|s_{t+\tau}) \!\!\prod_{\tau=-L+1}^0 \!\!p(s_{t+\tau}|s_{t+\tau-1}) p(s_{t-L}) \textrm{d}s_{t-L:t}
\end{align*}
We will first marginalize $s_{t-L}$. To do that, we notice that if $p(s_{t-L})$ follows a von Mises with mean $\hat{\mu}_{t-L}$ and concentration $\hat{\kappa}_{t-L}$, then we can write:
\begin{align*}
 & p(\hat{y}_{t-L}|s_{t-L})p(s_{t-L}) \\
 & = {\cal M}(\hat{y}_{t-L};s_{t-L},\hat{\omega}_{t-L}\kappa_y){\cal M}(s_{t-L};\hat{\mu}_{t-L},\hat{\kappa}_{t-L})\\
 &= {\cal M}(s_{t-L};\bar{\mu}_{t-L},\bar{\kappa}_{t-L}) \frac{I_0(\bar{\kappa}_{t-L})}{2\pi I_0(\hat{\omega}_{t-L}\kappa_y)I_0(\hat{\kappa}_{t-L})}
\end{align*}
with
\begin{align*} 
 \bar{\mu}_{t-L} &= \tan^{-1}\left( \frac{\hat{\omega}_{t-L}\kappa_y\sin\hat{y}_{t-L} + \hat{\kappa}_{t-L}\sin\hat{\mu}_{t-L}}{\hat{\omega}_{t-L}\kappa_y\cos\hat{y}_{t-L} + \hat{\kappa}_{t-L}\cos\hat{\mu}_{t-L}} \right),\\
 \bar{\kappa}_{t-L}^2 &= (\hat{\omega}_{t-L}\kappa_y)^2 + \hat{\kappa}_{t-L}^2 + 2\hat{\omega}_{t-L}\kappa_y\hat{\kappa}_{t-L}\cos(\hat{y}_{t-L}-\hat{\mu}_{t-L}),
\end{align*}
where we used the harmonic addition theorem.

Now we can effectively compute the marginalization. The two terms involving $s_{t-L}$ are:
\begin{align*}
 &\int {\cal M}(s_{t-L+1};s_{t-L},\kappa_d){\cal M}(s_{t-L};\bar{\mu}_{t-L},\bar{\kappa}_{t-L}) d s_{t-L} \\
 &\approx {\cal M}(s_{t-L+1};\hat{\mu}_{t-L+1},\hat{\kappa}_{t-L+1}) 
\end{align*}
with
\begin{align*}
 \hat{\mu}_{t-L+1} &= \bar{\mu}_{t-L}, \\
 \hat{\kappa}_{t-L+1} &= A^{-1}( A(\bar{\kappa}_{t-L}) A(\kappa_d) ).
\end{align*}

Therefore, the marginalization with respect to $s_{t-L}$ yields the following result:
\begin{align*}
 \int& p(\hat{y}_{t-L:t},s_{t-L:t}) \textrm{d}s_{t-L:t}\\
  &= \int\prod_{\tau=-L}^0 p(\hat{y}_{t+\tau}|s_{t+\tau}) \!\!\prod_{\tau=-L+1}^0 \!\!p(s_{t+\tau}|s_{t+\tau-1}) p(s_{t-L}) \textrm{d}s_{t-L:t}\\
  &= \frac{I_0(\bar{\kappa}_{t-L})}{2\pi I_0(\hat{\omega}_{t-L}\kappa_y)I_0(\hat{\kappa}_{t-L})}\int\prod_{\tau=-L+1}^0 p(\hat{y}_{t+\tau}|s_{t+\tau}) \times \\
  & \qquad\prod_{\tau=-L+2}^0 \!\!p(s_{t+\tau}|s_{t+\tau-1})p(s_{t-L+1}) \textrm{d}s_{t-L+1:t}.
\end{align*}

Since we have already seen that $p(s_{t-L+1})$ is also a von Mises distribution, we can use the same reasoning to marginalize with respecto to $s_{t-L+1}$. This strategy yields to the recursion presented in the main text.

\section{Results with errors}
\label{sec:appendix-m-step}

\begin{figure}[h]
\begin{center}
\begin{tabular}{cc}
\includegraphics[scale=0.25]{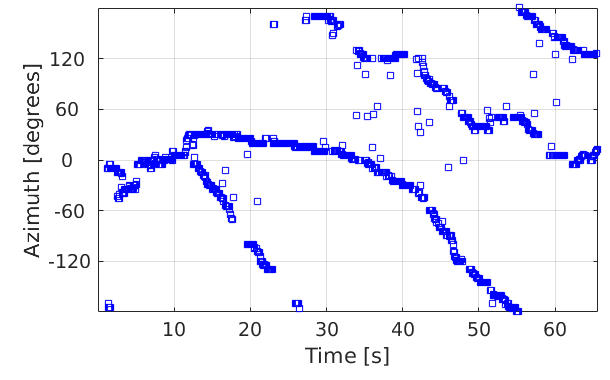} &
\includegraphics[scale=0.25]{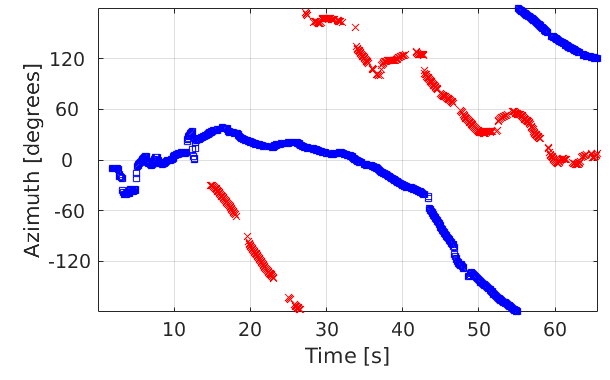} \\
(a) vM-PHD [13] & (b) GM-FO [15]\\
\includegraphics[scale=0.25]{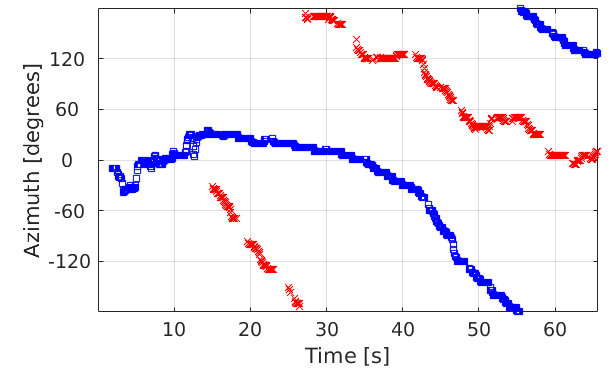} &\includegraphics[scale=0.25]{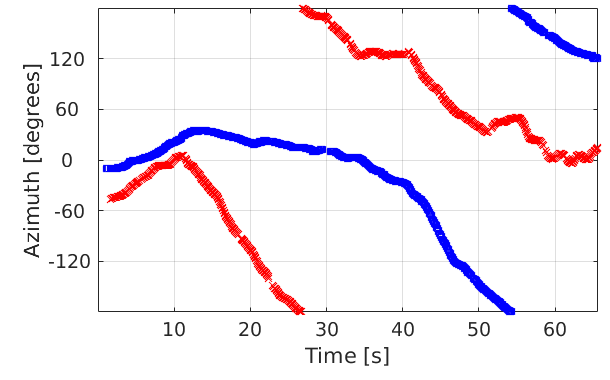} \\
(c) vM-VEM (proposed) & (d) ground-truth trajectories\\
\end{tabular}
\end{center}
\caption{Results obtained with recording \#3 from Task 6 of the LOCATA dataset. Different colors represent different audio sources. Note that vM-PHD is unable to associate sources with trajectories.}
\label{fig:additional-results}
\end{figure}

\bibliographystyle{IEEEtran}

\end{document}